\documentclass[twocolumn,superscriptaddress,showpacs,showkeys,amsmath,amssymb,aps]{revtex4}

\usepackage{epsfig}
\usepackage{graphicx}
\usepackage{bm}

\def\slashchar#1{\setbox0=\hbox{$#1$}
\dimen0=\wd0 \setbox1=\hbox{/} \dimen1=\wd1
\ifdim\dimen0>\dimen1 \rlap{\hbox to \dimen0{\hfil/\hfil}} #1
\else \rlap{\hbox to \dimen1{\hfil$#1$\hfil}} / \fi}

\begin{document}

\title{Update of the Hagedorn mass spectrum}

\author{Wojciech Broniowski}%
\email{Wojciech.Broniowski@ifj.edu.pl} 
\affiliation{The H. Niewodnicza\'nski Institute of Nuclear Physics,
Polish Academy of Sciences, PL-31342 Krak\'ow, Poland}
\author{Wojciech Florkowski}%
\email{florkows@amun.ifj.edu.pl} 
\affiliation{The H. Niewodnicza\'nski Institute of Nuclear Physics,
Polish Academy of Sciences, PL-31342 Krak\'ow, Poland}
\affiliation{Institute of Physics, \'Swi\c{e}tokrzyska Academy,
ul.~\'Swi\c{e}tokrzyska 15, PL-25406~Kielce, Poland}
\author{Leonid Ya. Glozman}%
\email{glozman@kfunigraz.ac.at}
\affiliation{Institute for Theoretical Physics, University of Graz,
Universit{\"a}tsplatz 5, A-8010 Graz, Austria}
\date{June 21, 2004}

\begin{abstract}
We present an update of the Hagedorn hypothesis of the exponential growth of the 
number of hadronic resonances with mass. We use the newest available experimental 
data for the non-strange mesons and baryons, as well as fill in some missing states 
according to the observation that the high-lying states form chiral multiplets. The 
results show, especially for the case of the mesons, that the Hagedorn growth 
continues with the increasing mass, with the new states lining up along the
exponential growth.
\end{abstract}

\pacs{25.14.20.-c, 14.40.-n, 12.40Yx, 12.40Nn}

\keywords{particle spectra, Hagedorn hypothesis, chiral symmetry}

\maketitle

The Hagedorn hypothesis \cite{h1,h2,h3} of the exponential growth of the number of 
hadronic resonances with mass is one of the most fundamental issues in particle 
physics. The formula for the asymptotic dependence of the density of hadronic states 
on mass, namely
\begin{eqnarray}
\rho(m) = f(m) \exp(m/T_H),
\label{hag}
\end{eqnarray}
where $f(m)$ denotes a slowly varying function and $T_H$ is the {\em Hagedorn temperature}, 
has gained a lot of attention due to its appealing simplicity, fundamental character, 
support from the experimental data and theoretical approaches, as well as because of its relevance 
to the phenomenology of particle production, in particular concerning the possible phase
transition from the hadron gas to the quark-gluon plasma
\cite{CabibboParisi,Chernavskaya:1994zj,Blanchard:2004du}.

The purpose of this note is to present an update of the experimental verification 
of Eq.~(\ref{hag}). We supplement the data published in the Particle Data Tables 
\cite{PDG} with the new experimental information \cite{b1,b5}, 
as well 
as add theoretically predicted new states belonging to chiral multiplets
\cite{g1,cg1,g2,g3,g4}. 
Although the appearance of some of these states has not been verified experimentally yet,
their existence follows from the recent theoretical findings that the high-lying 
particle spectrum essentially has the features of {\em restored chiral symmetry} \cite{g1}. 

The results are shown for the non-strange mesons and baryons, where the new data is 
available. The paper has no pretence of presenting new models or ideas; nevertheless, 
due to the fundamental nature of the problem related to basic ideas behind the 
formation of bound states and resonances in particle physics, the results of our 
simple compilation should be of interest for the community. We include the new 
experimental results and show that the new data are important in the verification of the
Eq.~(\ref{hag}). The new results extend significantly, at least for the mesons, the range of 
fiducial range of the Hagedorn hypothesis. While with the data listed in the 1998 edition 
of the Particle Data Tables \cite{PDG} used in \cite{twoT,Meson2000,Bled} the 
exponential growth for non-strange mesons could be observed up to the masses of about 
1.8~GeV, now it continues higher up, till about 2.3~GeV. 

We start with a very brief reminder of the history of the Hagedorn hypothesis 
(for much more complete historical presentations we refer the reader to Hagedorn's original 
lecture \cite{h3} and to a tribute article by Ericson and Rafelski \cite{Ericson}). 
Equation (\ref{hag}) was originally proposed to explain the spectra in the $p$-$p$ and 
$\pi$-$p$ scattering \cite{h1, HagRan}. Later, it was obtained from the statistical 
bootstrap models \cite{h2,Frautschi,Nahm,Yellin}. Subsequently, it gained a 
convincing support from the dual string models \cite{Huang,Cudell,Dienes,Jacob}. 
It is worthwhile to recall that in the 1960s, when the original Hagedorn idea was 
formed, very few hadronic states were known, up to the mass of the $\Delta$ isobar. More 
and more states have been accumulated over the years, thus much more systematic 
studies were possible, such as for instance the analysis of Ref.~\cite{Rafelski} and 
of Ref.~\cite{twoT}, where two of us (WB,WF) pointed out the different growth rate 
of mesons and baryons, as well as demonstrated the universality of the Hagedorn 
temperatures with strangeness. The faster growth of the baryon spectrum was also 
noted in Ref.~\cite{Freund}. 

The Hagedorn concept of the limiting temperature appears in many different
contexts, {\em e.g.}, in the studies of non-linear Regge trajectories
\cite{Brisudova:1998wq,Burakovsky:1998ct,Brisudova:1999ut}, strings
\cite{Atick:1988si,Sundborg:1999ue,Gubser:2000mf}, d-branes \cite{Abel:2000jq},
and cosmology \cite{Leblanc:1988eq}. Moreover, a complete treatment of hadronic 
resonances, as suggested by Hagedorn already in the 1960s, is the basic ingredient
of the successful models of hadron production in heavy-ion collisions at 
the RHIC energies \cite{Torrieri:2004zz,SA}.

After many dormant years with essentially no incoming new data, 
a recent systematic partial wave analysis of the $\bar p p$
annihilation at LEAR has revealed a lot of new meson states in the mass
range 1.8 - 2.4~GeV \cite{b1,b5}. These new experimental results
turned out to be in line with the proposed idea that the 
spontaneously broken chiral symmetry of QCD should be effectively
restored in the highly excited hadrons (one terms this phenomenon as the
{\em chiral symmetry
restoration of the second kind}) \cite{g1,cg1,g2}. This kind of
chiral symmetry restoration implies that the excited hadron states
fill out multiplets of the chiral $U(2)_L \times U(2)_R$ group.
Indeed, the newly discovered meson states \cite{b1,b5}
turned out to systematically fall into almost degenerate chiral
multiplets with a few missing states yet to be discovered 
\cite{g3,g4}.

In this note we extend the analysis of Refs.~\cite{twoT,Meson2000} and include
all mesons listed in Refs.~\cite{g3,g4}. We stress that in addition to the 
experimental states which have been reported in Refs.~\cite{b1,b5} we add a few
still missing states (marked with the question signs in Refs.~\cite{g3,g4})
and reconstruct their energies according to the known
energies of their chiral partners. We consider only the $J=0$, $1$, $2$, and $3$ states,
where the experimental information is rather complete. 

In addition to these states we also consider the states with hidden
strangeness, {\em i.e.} composed of the $\bar s s$ pairs. These states
could not be seen in $\bar p p$. Hence here our procedure is somewhat
more speculative. We assume that any isosinglet 
$\bar n n = \frac{\bar u u + \bar d d}{\sqrt 2}$, which is experimentally
seen in $\bar p p$, should be accompanied by an $\bar s s$ state
with the mass approximately 200~MeV higher. Hence, given the complete
amount of the $\bar n n$ states listed in Refs.~\cite{g3,g4} we add the corresponding
$\bar s s$ states.

\begin{figure}[tb]
\begin{center}
\epsfig{figure=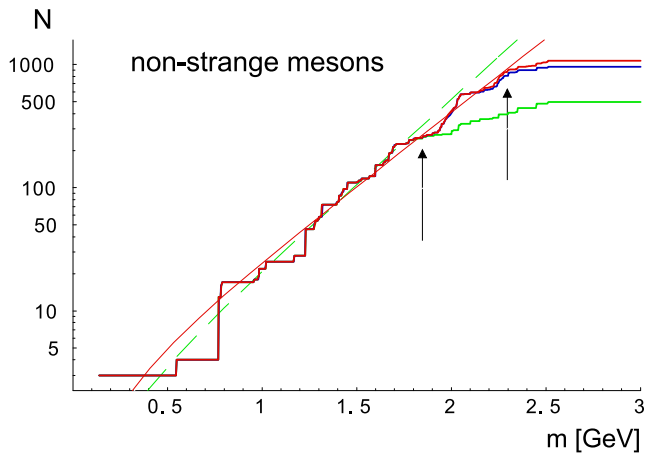,width=0.48\textwidth}
\end{center}
\caption{Accumulated spectrum of non-strange mesons plotted as a 
function of mass (step-like lines). 
The lower curve at high $m$ corresponds to particles listed in the Particle Data 
Tables of Ref.~\cite{PDG}, while the higher two curves include the new experimental 
and theoretical states as described in the text. The middle curve includes the
states listed in Refs.~\cite{g3,g4}, while the top curve adds the states with
hidden strangeness. The thin dashed (solid) line
corresponds to the exponential fit to the spectra of the old (new) data. 
The arrows indicate the approximate upper values in $m$ of the validity 
of the Hagedorn hypothesis for the old and new data, respectively.}
\label{fig:mes}
\end{figure}

Rather than comparing the density of states $\rho(m)$ itself to the data, it is 
customary to form the accumulated number of states of mass lower than $m$, 
%
\begin{equation}
N_{{\rm exp}}(m)=\sum_{i}g_{i}\Theta (m-m_{i}),
\label{accu}
\end{equation}
where $g_{i}=(2J_{i}+1)(2I_{i}+1)$ is the spin-isospin degeneracy of the $i$%
th state, and $m_{i}$ is its mass. The theoretical counterpart of Eq.~(\ref{accu}) is
\begin{equation}
N_{{\rm theor}}(m)=\int_{0}^{m}\rho(m^{\prime })dm^{\prime }.
\label{accuth}
\end{equation}
Working with $N(m)$ rather than $\rho(m)$ conveniently avoids the need of building 
histograms, but clearly it is a purely technical issue and the conclusions drawn 
below remain unchanged if one decides to work with $\rho(m)$ itself.

The results of our compilation for non-strange mesons are shown in Fig.~\ref{fig:mes}.
The lines with steps correspond to Eq.~(\ref{accu}). Above $m=1.8$~GeV the curves split
into three, with the lower one representing the compilation of Ref.~\cite{twoT} based of the 1998 
review of PDG \cite{PDG}. The middle curve contains in addition the states 
listed in refs.~\cite{g3,g4}, 
while the top curve includes also the hidden-strangeness states, as described above. 
It is clear from Fig.~\ref{fig:mes} that the included new states nicely 
line up along the exponential growth, thus extending the range of the Hagedorn hypothesis 
seen in the data. We also note that adding up the hidden-strangeness states 
has a much smaller effect than adding the states of Refs.~\cite{g3,g4}, which is simply due to 
a lower isospin degeneracy factor. 

The thin solid lines in Fig.~\ref{fig:mes} show the results of the exponential fits with $f(m)=1$ in Eq.~(\ref{hag},
\ref{accuth}), which is the simplest choice. While for the old data the least-squares method
yields $\rho(m)=2.84/{\rm GeV}
\exp[m/314~{\rm MeV})]$, with the states of Ref.~\cite{g3,g4} included we obtain 
$\rho(m)=4.73/{\rm GeV}
 \exp[m/(367~{\rm MeV})]$, and with the additional $\bar s s$
states we get $\rho(m)=4.52/{\rm GeV}
 \exp[m/362~{\rm MeV})]$. The fit was made up to $m=1.8$~GeV with the old data 
and up to $m=2.3$~GeV with the new data. The higher value for $T_H$ obtained with the new data corresponds 
to the lower slope in Fig.~\ref{fig:mes}. Certainly, the values of the fitted parameters should 
be taken with care, since they also reflect the assumed fitting range in $m$. It should also be 
noted, that adding more states in the range around 2~GeV, when experimentally found, 
would increase the slope, thus decreasing $T_H$. 

In this place the reader may be a bit surprized with the quoted high values of $T_H$, much higher than the 
typically cited values in the range of 200~MeV. The issue, as discussed in detail in Ref.~\cite{Bled}, 
has to do with the choice of the ``slowly-varying'' function $f(m)$. The point is that typical 
model predictions for this function are not so slowly varying in the range of data. For instance, 
with the original Hagedorn choice $f(m)={\rm const}/(m^2+500~{\rm MeV}^2)^{5/4}$ we get much 
lower values for $T_H$. With this form we obtain for the bottom to top curves of Fig.~\ref{fig:mes}
the following values:
$T_H=196$, $230$, and $228$~MeV, respectively. The choice of the fitting range in $m$ is 
as stated above.

\begin{figure}[tb]
\begin{center}
\epsfig{figure=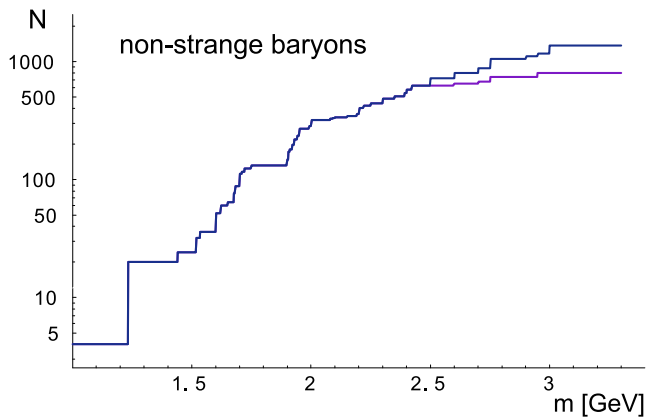,width=0.48\textwidth}
\end{center}
\caption{Accumulated spectrum of non-strange baryons plotted as a 
function of $m$. 
The lower curve at high $m$ corresponds to older data of Ref.~\cite{PDG}, while the 
higher curve includes the new states as described in the text. }
\label{fig:bar}
\end{figure}

Now we pass to the case of the non-strange baryons.
With the help of identification of states in chiral multiplets \cite{cg1},
we add the missing states (marked with the question signs in \cite{cg1})
on top of the states from PDG \cite{PDG} used in Ref.~\cite{twoT}.
In this way we fill the chiral multiplets. The results 
of this procedure are shown in Fig.~\ref{fig:bar}. 
We note that the effect of including these baryon states is less important than 
in the analogous procedure for the mesons. In the present case we do not show the 
fit to the exponential formula, since it is difficult to line-up the results along one straight
line in a sufficiently broad range of $m$. Actually, with the present data one may see
a straight line up to about $m=2$~GeV, and possibly another straight line, with a lower slope, above.
However, this may be an artifact of missing data in the high-mass range. 

Indeed, the parity doublets in $N$ and $\Delta$ can be associated with the
$(0,1/2) \oplus (1/2,0)$ representations for the nucleon spectrum,
the $(0,3/2) \oplus (3/2,0)$ multiplets in the $\Delta$ spectrum, and with
the $(1,1/2) \oplus (1/2,1)$ representations which combine the doublets
in the nucleon and delta spectra. If all these multiplets are realized in
nature, then the number of the states in the region above 2~GeV should be
much larger than given in PDG. Unfortunately, this region has never been systematically
explored in experiments.

\begin{figure}[tb]
\begin{center}
\epsfig{figure=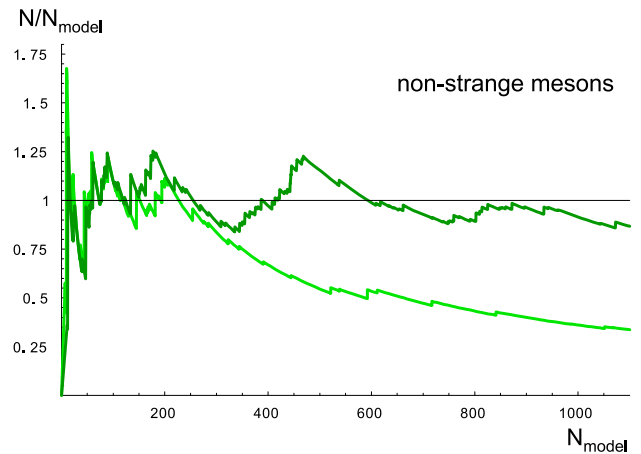,width=0.48\textwidth}
\end{center}
\caption{The ratio of the accumulated spectrum of non-strange baryons to the 
exponential fit, plotted as a function of $m$. 
The lower curve at high $m$ corresponds to older data of Ref.~\cite{PDG}, while the 
higher curve includes the new states as described in the text. 
We note a sizeable increase of the validity range of the Hagedorn hypothesis.}
\label{fig:mesrat}
\end{figure}

We now come back to the meson case of Fig.~\ref{fig:mes}, and wish to present the data in a 
somewhat different manner. The problem of the presentation in the $\log$ scale, as in 
Fig.~\ref{fig:mes}, is that the low-mass states are sparse, while the high-mass states are 
jammed up. For that reason we now look at the ratio of the experimental function (\ref{accu}) 
to the model function (\ref{accuth}), with the choice $f(m)=1$ and the parameters at the fitted 
values quoted in the text. The ratio is plotted as a function of the accumulated number of model states, 
$N_{\rm model}$. If the Hagedorn hypothesis complies to the data, this ratio should be
equal to unity. Indeed, this is so with the new data up to about 900 states, while with the 
old data it was true only up to about 250 states. Again, we see vividly that the inclusion of the 
new states significantly increases the range of validity, or verification, of Eq.~(\ref{hag}).

\begin{figure}[tb]
\begin{center}
\epsfig{figure=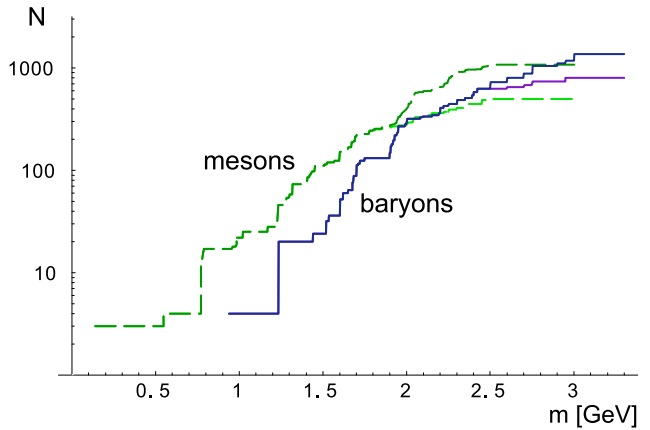,width=0.48\textwidth}
\end{center}
\caption{Comparison of mesons (dashed lines) and baryons (solid lines) of 
Figs.~\ref{fig:mes} and \ref{fig:bar}.}
\label{fig:mesbar}
\end{figure}

Finally, for the reader's convenience we overlay our results for the mesons and baryons in 
one plot of Fig.~\ref{fig:mesbar}. As pointed out in Ref.~\cite{twoT}, up to $m=2$~GeV we note a faster growth 
rate for baryons than for mesons, which means two distinct Hagedorn temperatures for mesons and 
baryons. This is a prediction of dual string models, see Ref.~\cite{Bled} for a discussion. 
For higher masses this feature is no longer obvious, with more experimental information needed 
to clarify the issue.

In conclusion, we list our main observations:

\begin{enumerate}
\item The newly-observed meson states lead to a continued exponential growth on the number of states
with mass, in accordance to the Hagedorn hypothesis, which now reaches up to masses of about 2.3~GeV.
\item For the baryons the situation is less clear, with the exponential growth seen up to about 2~GeV. 
\item The inclusion of the missing states based on the identification of chiral multiplets 
helps to comply to the Hagedorn hypothesis at high masses. 
\item Certainly, more experimental data in the high-mass range are highly desired 
to investigate further and with greater detail the hadron spectroscopy. 
\end{enumerate}

\begin{acknowledgments}
The authors wish to thank the organizers of the {\em XLIV Cracow School of Theoretical Physics} in Zakopane 
and the {\em Mini-Workshop Bled 2004: Quark Dynamics}, where this research was completed.
WB and WF acknowledge the support of the Polish State Committee of Scientific Research, grant 
2~P03B~05925. LG acknowledges the support of the FWF project P16823-N08 of the Austrian Science Fund.

\end{acknowledgments}

\end{document}